\documentclass[twocolumn,preprintnumbers,amsmath,amssymb,amsfonts,prl,longbibliography]{revtex4-1}

\usepackage{amsmath}
\usepackage{graphicx}
\usepackage[export]{adjustbox}
\usepackage{dcolumn}
\usepackage{bbm}
\usepackage{bm}
\usepackage{subfigure}
\usepackage{amssymb}
\usepackage{hyperref}
[    linkcolor=black,urlcolor=black]
\hypersetup{
    colorlinks=true,
    citecolor=blue,
    }
\usepackage{xcolor}
\usepackage{subfigure}
\usepackage{makecell}
\usepackage{mathtools}
\usepackage{booktabs}
\usepackage{xparse}
\usepackage{wasysym}
\usepackage{tikz-feynman}
\usepackage{color}
\usetikzlibrary{decorations.pathreplacing,decorations.markings}

\usepackage{soul}
\usepackage{color}
\usepackage[normalem]{ulem}
\usepackage{comment}

\definecolor{darkblue}{rgb}{0.,0.,0.4}
\definecolor{darkred}{rgb}{0.5,0.,0.}
\definecolor{BlueViolet}{RGB}{138,43,226}
\definecolor{SkyBlue}{RGB}{30,144,255}
\definecolor{DarkGreen}{RGB}{0,100,0}

\stepcounter{secnumdepth}
\stepcounter{tocdepth}

\begin{document}

\title{$\mathbb Z_2$ spin liquids in the higher spin-$S$ Kitaev honeycomb model: \\ An exact deconfined $\mathbb Z_2$ gauge structure in a non-integrable model}

\author{Han~Ma}
\affiliation{\vspace*{0.3 cm}Perimeter Institute for Theoretical Physics, Waterloo, Ontario N2L 2Y5, Canada}

\begin{abstract}
The higher spin Kitaev model prominently features the extensive locally conserved quantities the same as the spin-$1/2$ Kitaev honeycomb model, although it is not exactly solvable. 
It remains an open question regarding the physical meaning of these conserved quantities in the higher spin model.
In this Letter, by introducing a Majorana parton construction for a general spin-$S$ we uncover that these conserved quantities are exactly the $\mathbb Z_2$ gauge fluxes in the general spin-$S$ model, including the case of spin-$1/2$.
Particularly, we find an even-odd effect that the $\mathbb Z_2$ gauge charges are fermions in the half integer spin model, but are bosons in the integer spin model.
We further prove that the fermionic $\mathbb Z_2$ gauge charges are always deconfined; hence the half integer spin Kitaev model would have non-trivial spin liquid ground states regardless of interaction strengths in the Hamiltonian.
The bosonic $\mathbb Z_2$ gauge charges of the integer spin model, on the other hand, could condense, leading to a trivial product state, and this is indeed the case at the anisotropic limit of the model.

\end{abstract}

\maketitle

Frustrated magnets exhibit novel behaviors and are platforms for highly quantum entangled phases, such as spin liquids~\cite{Savary2017quantum}.  
So far, searching for the spin liquid phases in experiments remains a challenging task. On one hand, experimentally relevant models are, in general strongly interacting and hard to solve, making their ground states mysterious and controversial. 
On the other hand, spin liquids are established in the exactly solvable models that, however, are usually unrealistic for experiments.
In these regards, the spin-$1/2$ Kitaev honeycomb model~\cite{kitaev2006anyons} attracts lots of interest due to its exact solvability, its realization of distinct spin liquid phases, and its relevance to materials such as $\alpha-{\rm RuCl}_3$ and ${\rm A}_2{\rm IrO}_3$~\cite{Jackeli2009mott,rau2016spin,hermanns2018physics,trebst2022kitaev}.

The spin-$1/2$ Kitaev model was originally solved by the parton construction, an approach widely used in the study of spin liquid physics in many different systems.
As is well known, the parton construction is uncontrolled and biased for most of the systems, but it is exact for the spin-$1/2$ Kitaev model, which can be attributed to the existence of extensive local conserved $\mathbbm{Z}_2$ fluxes.
The corresponding $\mathbbm{Z}_2$ gauge structure is also manifest in the exact parton construction.
It remains an outstanding question that if such an exact parton construction can be generalized to other theoretical models, in particular, to the models that are not exactly solvable yet experimentally relevant. 

\begin{figure}[tb]
    \centering
    \includegraphics[width=.18\textwidth]{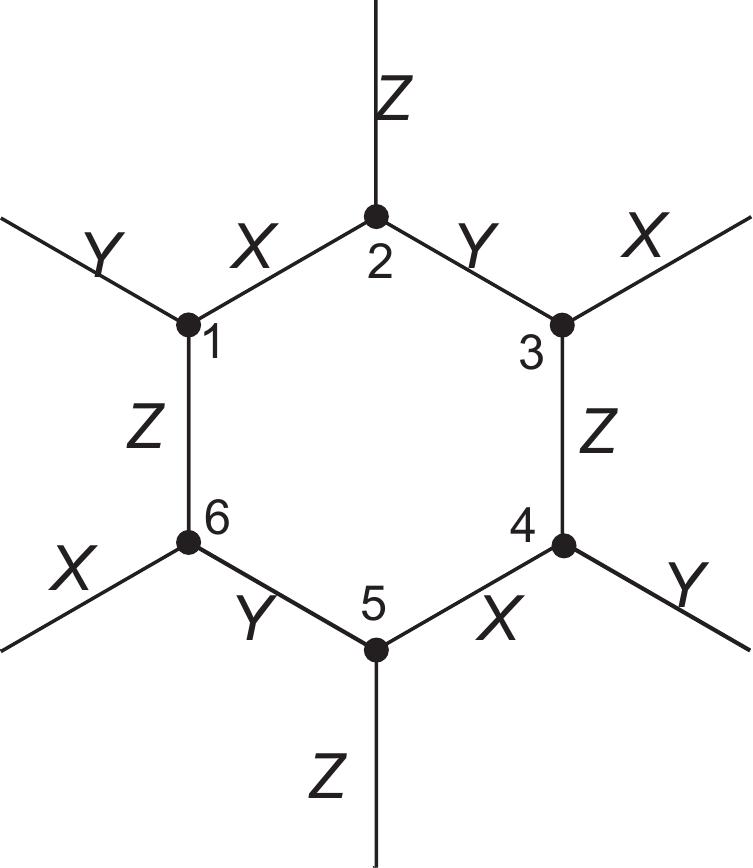}
    \caption{The $x$, $y$ and $z$ bonds of the honeycomb lattice. The numbers label the sites around a hexagonal plaquette.}
    \label{fig:honeycomb}
\end{figure}

In this Letter, we provide a positive answer to the intriguing question of the exact parton construction in a non-integrable model. 
We consider the generalized higher spin-$S$ Kitaev honeycomb model,  
\begin{eqnarray}
    H= - \sum_\mu J_\mu \sum_{\langle ij\rangle \in \mu} S^\mu_i S^\mu_j .
    \label{eq:H_S_Kitaev}
\end{eqnarray}
As in the spin-$1/2$ Kitaev model, the spin-spin interaction pointing a certain direction is locked to one of the bond orientations $\mu=x,y,z$ which are demonstrated in Fig.~\ref{fig:honeycomb}. Subscripts $i$ and $j$ denote the two ends of the bond $\mu$. 
The only difference is that now we consider a general spin-$S$ rather than spin-$1/2$.
Recently, due to its potential experimental realization~\cite{xu2018interplay,stavropoulos2019microscopic,xu2020possible,stavropoulos2021magnetic,samarakoon2021static,cen2022determining}, this higher spin Kitaev model generates a lot of interest and has been studied using various analytical and numerical tools~\cite{baskaran2008spin,oitmaa2018incipient,koga2018ground,Minakawa2019magnetic,zhu2020magnetic,hickey2020field,dong2020spin,Lee2020tensor,lee2021anisotropy,Khait2021characterizing,jin2022unveiling,chen2022excitation,Bradley2022instabilities,gordon2022insights, chen2022phase,minakawa2019quantum}.

In general, the spin-$S$ Kitaev model has exactly conserved $\mathbb{Z}_2$ valued local plaquette operators,
\begin{eqnarray}
    W_p (\varhexagon)= e^{i\pi S^y_{1}}~e^{i\pi S^z_{2}}~e^{i\pi S^x_{3}}~e^{i\pi S^y_{4}}~e^{i\pi S^z_{5}}~e^{i\pi S^x_{6}}
    \label{eq:W_S}
\end{eqnarray}
which take the product of $e^{i\pi S^\mu_i}$ around a hexagonal plaquette~\cite{baskaran2008spin}. 
These conserved quantities are the $\mathbb{Z}_2$ gauge fluxes in the spin-$1/2$ model and they are the key to the integrability of the spin-$1/2$ model.
The number of these conserved quantities, however, is not enough to make the higher spin model integrable due to its larger Hilbert space.
Nevertheless, by introducing a parton construction with $8S$ Majorana fermions, we show exactly that these conserved quantities are still $\mathbbm{Z}_2$ gauge fluxes.
More interestingly, we have found an even-odd effect, namely the $\mathbbm{Z}_2$ gauge charge is a fermion or boson in the half integer or integer spin model, respectively.
Moreover, we prove that the fermionic $\mathbbm{Z}_2$ gauge charge is always deconfined in the half integer spin model.
In other words, the half integer spin model always has a spin liquid ground state no matter what the interaction strengths $J_{x,y,z}$ are.
In contrast, for the integer spin model the bosonic $\mathbb{Z}_2$ gauge charge could condense, giving rise to a trivial paramagnet, and it is exactly the case in the anisotropic limit, e.g. $J_z\gg J_{x,y}$.

\begin{widetext}
\begin{center}
\begin{table}[h]
    \centering
    {\setcellgapes{1.2ex}\makegapedcells
    \begin{tabular}{c|c|c|c|c} \hline 
        Representations   & Hilbert space & Operator  & Constraints & Physical states\\ \hline \hline
        spin-$S$              & $2S+1$ & $S^\mu$ &  - & $|\{S^z\}\rangle $\\ \hline
        $2S$ spin-$1/2$        & $2^{2S}$ & $S^\mu = \sum_{a=1}^{2S} S_a^\mu $ &  $\mathcal{C}_s: \sum_\mu (\sum_{a=1}^{2S} S^\mu_a)^2 = S(S+1)$ & $\mathcal{P}_s|\{S^z_a\}\rangle $\\\hline
        $8S$ Majorana fermions & $4^{2S}$ & \makecell{$2S_a^\mu=i\gamma_a^0\gamma^\mu_a$\\\vspace{-3mm} \\ $2S^\mu = i\sum_{a=1}^{2S} \gamma_a^0 \gamma^\mu_a $ }& \makecell{$\mathcal{C}_s:\sum_\mu (\sum_{a=1}^{2S} \gamma^0_a\gamma^\mu_a)^2 = -S(S+1)$ \\\vspace{-3mm} \\  $\mathcal{C}_\gamma^a: \gamma^0_a \gamma^x_a\gamma^y_a\gamma^z_a =1$ }&\makecell{
        $\mathcal{P}_s \prod_{a=1}^{2S}\mathcal{P}_{\gamma}^a|\{\gamma\}\rangle $}\\\hline
    \end{tabular}}
    \caption{Summary of different representations of a spin-$S$ operator. The spin operator $S^\mu$ with $\mu=x,y,z$ can be expressed in terms of the spin-$1/2$ operators or Majorana fermions that enlarge the original local Hilbert space. Physical states can be obtained by acting projection operator $\mathcal{P}_{s,\gamma}$, or equivalently imposing local constraints $\mathcal{C}_{s,\gamma}$ on the parton states . }
    \label{tab:summary_parton}
\end{table}
\end{center}
\end{widetext}

\emph{Parton construction.--} We start with reviewing the Majorana fermion representation of a spin-$1/2$. In terms of four Majorana fermions denoted as $\gamma^{0,x,y,z}$ satisfying relations $\gamma^\alpha\gamma^\beta+ \gamma^\beta\gamma^\alpha=2\delta_{\alpha\beta}$ for $\alpha,\beta=0,x,y,z$, spin-$1/2$ operators $S^\mu$ with $\mu=x,y,z$ can be expressed as 
$2S^\mu =i\gamma^0\gamma^\mu
$.
One can check the operators on the right side obey the commutation relations of the spin-$1/2$ operators $\left[S^\mu,S^\nu\right]=i \varepsilon_{\mu\nu\lambda}S^\lambda$
and $2\{S^\mu,S^\nu\}=\delta_{\mu\nu}I$. 
Meanwhile the identity $-8iS^x S^y S^z =I$ gives a local constraint among the four Majorana fermions as $\mathcal{C}_\gamma:~\gamma^0\gamma^x\gamma^y\gamma^z=1$. In the representation of Majorana fermions, the original two dimensional Hilbert space of a local spin is enlarged to a four dimensional Fock space of two complex fermions that may be defined as $c^\dag=\gamma^0+i\gamma^z$ and $d^\dag=\gamma^x+i\gamma^y$.
Accordingly, a physical state $|\Phi_{\rm phys}\rangle$ of the spins corresponds to the state of Majorana fermions $|\gamma\rangle$ satisfying the local constraint $\mathcal{C}_\gamma$. To put it another way, $|\Phi_{\rm phys}\rangle=\mathcal{P}_\gamma|\gamma\rangle$ where $\mathcal{P}_\gamma= \frac{\gamma^0\gamma^x\gamma^y\gamma^z+1}{2}$. In terms of these four Majorana fermions, the $\mathbbm{Z}_2$ gauge structure is manifest in the Hamiltonian of the Kitaev honeycomb model which can be rewritten as a fermionic charge $\gamma^0$ coupled to a $\mathbbm{Z}_2$ gauge field.

In general, we can have a similar parton construction for a spin-$S$. In terms of $8S$ Majorana fermions, a spin-$S$ operator can be represented as
\begin{eqnarray}
    2S^\mu =i \sum_{a=1}^{2S} \gamma^0_a \gamma^\mu_a =i \left[\gamma_1^0,\gamma_2^0,\dots, \gamma^0_{2S}\right]\left[\begin{array}{c}
        \gamma_1^\mu \\ \gamma_2^\mu \\ \vdots \\ \gamma^\mu_{2S}  
   \end{array}\right] , \label{eq:parton}
\end{eqnarray}
where $a$ labels the flavor of Majorana fermions ranging from $1$ to $2S$.
As usual, the Majorana fermions satisfy relations $\gamma^\alpha_a\gamma^\beta_b+ \gamma^\beta_a\gamma^\alpha_b=2\delta_{\alpha\beta}\delta_{ab}$. Accordingly, we can check that Eq.~\eqref{eq:parton} reproduces the commutation relation of the spin-$S$ operators. The dimension of the local Hilbert space is enlarged from $2S+1$ to $4^{2S}$. 
There are local constraints $\mathcal{C}_\gamma^a: \gamma^0_a\gamma^x_a\gamma^y_a\gamma^z_a=1$ for any $a \in [1,2S]$ and $\mathcal{C}_s: -\sum_\mu (\sum_{a,b=1}^{2S}\gamma^0_a \gamma^\mu_a \gamma^\mu_b \gamma^0_b)=4S(S+1)$, by imposing which the enlarged Hilbert space can be projected onto the original one.

This parton construction can be understood in the following two steps. First, a spin-$S$ operator can be decomposed to $2S$ spin-$1/2$ operators, i.e. $S^\mu =\sum_{a=1}^{2S}S^\mu_a$. This enlarges the local Hilbert space from $2S+1$ dimensions to $2^{2S}$ dimensions. Returning to the original Hilbert space requires the local constraint $\mathcal{C}_s: \sum_\mu (\sum_{a=1}^{2S} S^\mu_a)^2 = S(S+1)$. When $S=1/2$, this step is trivial and the constraint reduces to a trivial identity. Second, each spin-$1/2$ can be represented by four Majorana fermions as we demonstrated before. Together, the two steps summarized in Tab.~\ref{tab:summary_parton} yield Eq.~\eqref{eq:parton}.
This is actually the $USp(4S)$ parton construction studied before\cite{wen1999projective,barkeshli2010effective}, which is briefly reviewed in the supplemental material\cite{sm}.

Notice there is an explicit $O(2S)$ gauge redundancy in this parton construction, of which $(\gamma_{1}^\alpha\cdots \gamma_{2S}^\alpha)$ with $\alpha=0,x,y,z$ form $O(2S)$ vectors.
The physical spin is invariant under the $O(2S)$ rotation of the $8S$ Majorana fermions.
For $S=1/2$, the gauge group reduces to $O(1)\simeq \mathbbm{Z}_2$. For a system of higher spins, there is always a $\mathbbm{Z}_2$ subgroup which turns out to be important in the Majorana representation of the higher spin Kitaev model.
We can construct the charge of this $\mathbbm{Z}_2$ group as
\begin{eqnarray}
   \Gamma^{\alpha}&=& \frac{i^{S(2S-1)}}{(2S)!} \epsilon_{a_1,a_2,\dots,a_{2S}}\gamma_{a_1}^{\alpha}\gamma_{a_2}^{\alpha}\dots \gamma_{a_{2S}}^{\alpha} \nonumber\\
   &=&i^{S(2S-1)}\prod_{a=1}^{2S}\gamma_a^{\alpha}, \textrm{ with } \alpha=0,x,y,z ~,
   \label{eq:giant_Majorana}
\end{eqnarray}
where the repeated subscripts are summed over. 
It is an $SO(2S)$ singlet but is odd under the improper $\mathbbm{Z}_2$ transformation of the $O(2S)$. Hereafter, we call such an operator giant parton.
The prefactor ensures $( \Gamma^{\alpha}_i)^2=1$. 
This operator is a fermionic operator when $2S$ is odd, dubbed giant Majorana fermions. While for even $2S$, it is bosonic. As we will discuss below, it is the key to the different phases of matter in the integer and half-integer spin Kitaev honeycomb model.

\emph{Spin-S Kitaev honeycomb model.--} 
Using our parton construction in Eq.~\eqref{eq:parton}, the Hamiltonian in Eq.~\eqref{eq:H_S_Kitaev} can be expressed as 
\begin{eqnarray}
    H= - \sum_\mu J_\mu \sum_{\langle ij\rangle \in \mu}\sum_{a,b=1}^{2S} \gamma^0_{a,i}\gamma^\mu_{a,i}\gamma^\mu_{b,j}\gamma^0_{b,j}.
    \label{eq:H_S}
\end{eqnarray}
The local conserved plaquette operator can be written as a product of giant partons in Eq.~\eqref{eq:giant_Majorana},
\begin{eqnarray}
  W_p (\varhexagon)= \prod_{i,\mu \in \varhexagon}  \Gamma^\mu_{i}\Gamma^\mu_{i+\mu}
  =   \prod_{i,\mu \in \varhexagon} u^\mu_{i,i+\mu},
  \label{eq:flux_hexagon}
\end{eqnarray}
because $e^{i\pi S^\mu_j} = i^{2S} \prod_a \gamma_{a,j}^0 \gamma_{a,j}^\mu = i^{4S^2}\Gamma^0_j\Gamma^\mu_j$.
We can also define the operator on the $\mu$-bond as $u^\mu_{i,i+\mu}= \Gamma^\mu_{i}\Gamma^\mu_{i+\mu}$, which commute with each other as well as the Hamiltonian. 
Therefore, like the spin-$1/2$ Kitaev model,  $u^\mu_{i,i+\mu}$ can be viewed as a $\mathbbm{Z}_2$ gauge field and $W_p$ is the corresponding $\mathbbm{Z}_2$ flux.

\begin{figure}[h]
    \centering
    \includegraphics[width=.45\textwidth]{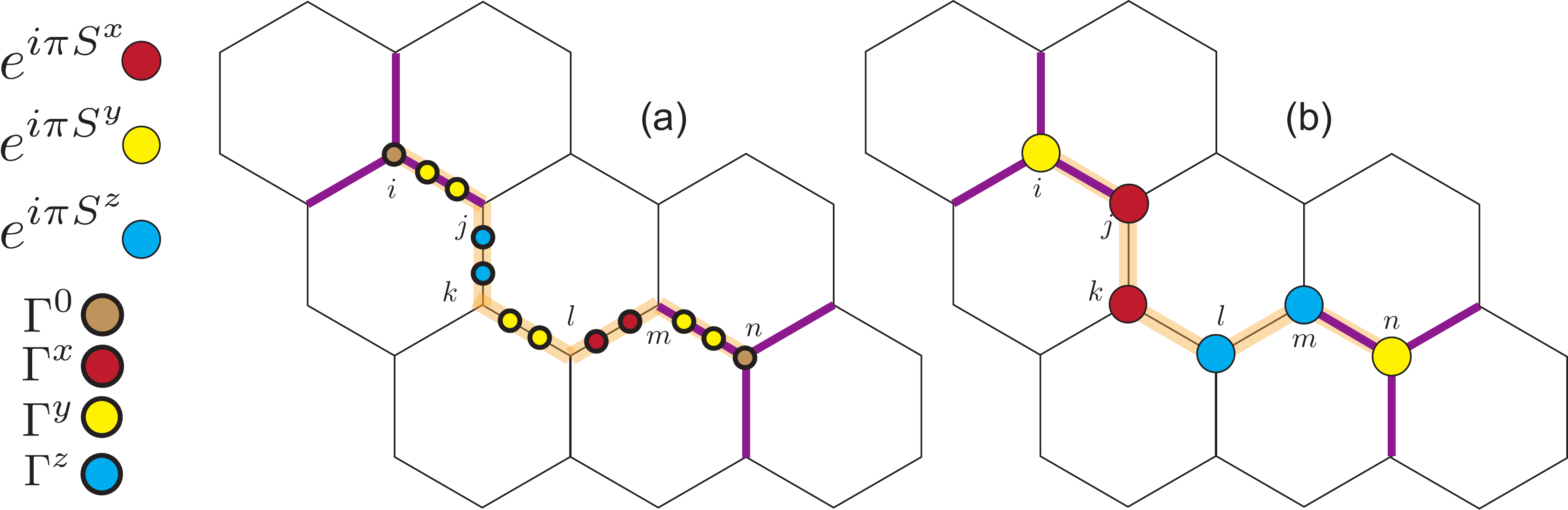}
    \caption{(a) and (b) are the string operators as a product of giant partons and a product of operators $e^{i\pi S^\mu}$, respectively, along the orange thick strings. Except for the three purple bonds intersecting at the two end points, the string operator commutes with the Hamiltonian. This indicates that the energy cost is only at the two end points and the string is tensionless. }
    \label{fig:excitations}
\end{figure}

Having identified the exact correspondence between local conserved quantities $W_p$ and the $\mathbb Z_2$ gauge fluxes, a natural question is whether this $\mathbb Z_2$ gauge field is deconfined or not. 
To answer this question, we can use the original definition of (de)confinement, namely to investigate how the energy cost of separating two gauge charges scales with their distance.
We can create and separate two $\mathbb Z_2$ gauge charges using a string operator, for example,
\begin{eqnarray}
   U
   &=& \Gamma^0_i u^y_{i,j} u^z_{j,k} u^y_{k,l} u^x_{l,m} u^y_{m,n}  \Gamma^0_{n} ,
   \label{eq:U_giant}
\end{eqnarray}
as shown in Fig.~\ref{fig:excitations}(a).
The giant partons $\Gamma^0_i$ and $\Gamma^0_n$ create  gauge charges at two end points of the string. The gauge fields $u$ connect two gauge charges making the string operator gauge invariant. 
This operator can be written in terms of physical spin operators,
\begin{eqnarray}
   U = e^{i\pi S^y_i}e^{i\pi S^x_{j}} e^{i\pi S^x_{k}} e^{i\pi S^z_{l}} e^{i\pi S^z_{m}} e^{i\pi S^y_{n}},
\end{eqnarray}
as shown in Fig.~\ref{fig:excitations}(b).

It is easy to verify that the Hamiltonian Eq.~\eqref{eq:H_S} commutes with the string operator $U$ in Eq.~\eqref{eq:U_giant} except at its end points. 
So the string is tensionless, or in other words, its energy cost would only come from its end points (i.e. gauge charges) and will not grow with the string length.
This is the familiar definition of deconfinement. 
However, there is one caveat, namely the condensation of gauge charge $\Gamma^0$ could also give rise to a tensionless string.
Specifically, the condensation of $\Gamma^0$ means $\langle \Gamma^0 \rangle \neq 0$, or more precisely,
\begin{equation}
\langle \Gamma^0_i (\prod_{\mu} u^{\alpha_\mu}_{\mu})   \Gamma^0_{j} \rangle = \textrm{const.}, \quad \textrm{for $|i-j|\gg 1$.}
\label{eq:condense_charge}
\end{equation}
In this case, the string operator does not create excitations, i.e., $\Gamma^0_i (\prod_{\mu} u^{\alpha_\mu}_{\mu})   \Gamma^0_{j} |gs\rangle = |gs\rangle$ (up to a phase factor).

In total, to have a deconfined $\mathbb Z_2$ gauge field, we need the tensionless string operator to create non-trivial gauge charge excitations.
For half integer $S$, the gauge charges $\Gamma_0$  are fermionic, so the string operator always creates non-trivial excitations. 
In other words, we prove that the half-integer spin Kitaev model with any coupling strength ($J_{x,y,z}$) is a spin liquid phase with deconfined fermionic spinon coupled to a $\mathbb Z_2$ gauge field. It is possible, although unlikely, that the entire gauge group is larger than $\mathbb Z_2$. As a result, besides the $\mathbb Z_2$ charge and flux, the spin liquid phase would possess other anyonic excitations.
For integer $S$, the  gauge charges $\Gamma_0$  are bosonic which can condense, leading to a trivial phase of matter. 
Next, we will show these bosons indeed condense in the integer spin Kitaev model in the anisotropic limit.

\emph{Anisotropic limit.--} 
Higher spin Kitaev model in the anisotropic limit (e.g., $J_z\gg J_x, J_y$) can be solved perturbatively, and its ground state is the $\mathbbm{Z}_2$ topological order for the half integer spin models or is a trivial product state in the integer spin models~\cite{kitaev2006anyons,Minakawa2019magnetic,lee2021anisotropy,minakawa2019quantum}.
This even-odd effect is a manifestation of the previously introduced giant parton, also known as a $\mathbb Z_2$ gauge charge, being a fermion or a boson, which we will explain in the degenerate perturbation theory\cite{sm} using our parton construction.

In the classical limit $J_{x,y}=0$, the ground state lives in an extensive degenerate manifold, with each $z$ bond hosting two-fold degenerate states $|S,S\rangle$ and $|-S,-S\rangle$ in the $S^z$ basis. 
Next, finite but small $J_{x,y}$ interactions would cause quantum tunneling between different classically degenerate states, leading to a new ground state of an effective quantum Hamiltonian in the anisotropic limit.

To derive this effective Hamiltonian in terms of Majorana partons, we first note that 
the partons are constrained by
$ \gamma^0_{a,i}\gamma^z_{a,i}\gamma^z_{b,\bar{i}}\gamma^0_{b,\bar{i}}=1$ for any $a,b \in [1,2S]$ in the space of ground states .
Then, 
within this space of degenerate ground states, $J_x$ and $J_y$ terms induce 
the effective Hamiltonian which turns out to be expressed merely as a product of the giant partons.

For the system of half-integer spins, a non-trivial effective Hamiltonian is obtained at the $8S$-th order of the perturbation
\begin{eqnarray}
   H_{\rm eff}^{\mathbbm{Z}+1/2} = -J_{\rm eff} \sum_{\diamond}  
   \includegraphics[width=3.8cm,valign=c]{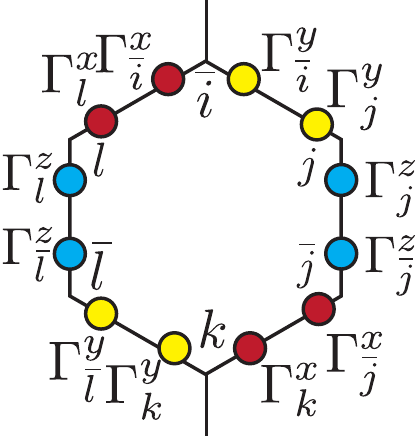}
   ,
   \label{eq:H_WP}
\end{eqnarray} 
where $J_{\rm eff} \propto J_x^{4S}J_y^{4S}/J_z^{8S-1} >0$. Each plaquette term is a product of $12$ giant Majorana fermions around a plaquette that is 
exactly the flux terms $W_p$ in Eq.~\eqref{eq:flux_hexagon} as well as Eq.~\eqref{eq:W_S}.
This is the familiar Wen-Plaquette model, or equivalently toric code model, which hosts the gapped $\mathbbm{Z}_2$ topological order phase~\cite{wen2003quantum,Kitaev2003fault}.
The giant fermionic parton $\Gamma^0$ is gapped in the anisotropic limit, and it is the deconfined fermionic excitation $\varepsilon$ in the $\mathbb Z_2$ topological order.

For the system of integer spins, we can obtain a non-trivial Hamiltonian at the $4S$-th order
\begin{eqnarray}
   H_{\rm eff}^{\mathbbm{Z}} &=& -J'_{\rm eff} \sum_z\Gamma^0_{l} \Gamma^x_{l}\Gamma^y_{\bar{l}}\Gamma^0_{\bar{l}} \nonumber\\
   &=& -J'_{\rm eff} \sum_z ~e^{i\pi S^x_l} e^{i\pi S^y_{\bar{l}}}  ,
\end{eqnarray}
where $J'_{\rm eff} \propto J_x^{2S}J_{y}^{2S}/J_z^{4S-1} >0$ and the sum is over all the $z$ bonds.
Notice the operator on the right hand side has $\mathbbm{Z}_2$ eigenvalues.
The ground state of this Hamiltonian is a trivial paramagnet corresponding to $\langle e^{i\pi S^x_l} e^{i\pi S^y_{\bar{l}}} \rangle =1$. We can check that Eq.~\eqref{eq:condense_charge} is satisfied in this trivial ground state since any string operator commutes with the Hamiltonian, indicating that the bosonic charge $\Gamma^0$ is condensed.

\emph{Isotropic limit.--} 
Finally, we will discuss the isotropic limit $J_x\sim J_y \sim J_z$. 
As we have proven before, the half integer spin model always has a deconfined $\mathbb Z_2$ gauge field coupled to fermionic spinons (i.e., giant Majorana fermions). 
The nature of the spin liquid further depends on whether the fermionic spinon is gapped or gapless, as summarized in Table.~\ref{tab:possible_phases_isotropic}. As we know, the isotropic spin-$1/2$ model is in a gapless phase with two Majorana cones at $\pm K$ points\cite{kitaev2006anyons}. So is the isotropic spin-$3/2$ model evidenced by numerics \cite{jin2022unveiling}. Therefore, 
we conjecture that any half-integer spin model is in a gapless phase and has two gapless giant Majorana cones at the $\pm K$ point of the Brillouin zone. It is stable under symmetric perturbations.
Possible phases of the integer spin model are also listed in Table.~\ref{tab:possible_phases_isotropic}, and a recent numerical work~\cite{Lee2020tensor} found that the isotropic limit of the spin-$1$ Kitaev model has a $\mathbb Z_2$ gapped spin liquid ground state whose gauge charge is a boson~\cite{chen2022excitation}.
Furthermore, we conjecture the $SO(2S)$ gauge field would be confined in both half integer and integer spin models, since there is no compelling reason for this large continuous gauge group to be deconfined.
\begin{table}[tb]
    \centering
    {\setcellgapes{1.2ex}\makegapedcells
    \begin{tabular}{c|c|c}\hline
         & $S=\mathbbm{Z}+1/2$ & $S=\mathbbm{Z}$ \\ \hline \hline
       Phase 1  & \makecell{ Gapless fermionic $\Gamma^0$ \\coupled to $\mathbbm{Z}_2$ gauge field }& \makecell{ trivial phase with \\ bosonic $\Gamma^0$ condensation}\\ \hline
       Phase 2  & \makecell{ Gapped fermionic $\Gamma^0$ \\coupled to $\mathbbm{Z}_2$ gauge field }& \makecell{Gapped bosonic $\Gamma^0$ \\coupled to $\mathbbm{Z}_2$ gauge field}\\\hline
    \end{tabular}}
    \caption{Possible phases of the isotropic higher spin Kitaev model.}
    \label{tab:possible_phases_isotropic}
\end{table}

\emph{ Conclusion.--} 
We introduce a generic Majorana parton representation for a higher spin-$S$ in the study of the higher spin Kitaev model, and it enables an exact identification of the extensive local commuting plaquette operators as conserved $\mathbbm{Z}_2$ gauge fluxes, similar to the case in the spin-$1/2$ Kitaev model.
Even though the higher spin model is not exactly solvable, we prove that the half integer spin models always have a deconfined $\mathbb Z_2$ gauge field coupled to fermionic spinons. 
In contrast, the integer spin model has the $\mathbb Z_2$ gauge field coupled to bosonic charges, which could condense and result in a trivial paramagnet, which is indeed the case in the anisotropic limit.
This even-odd effect of the integer and half-integer spin systems is a purely quantum effect, and it is reminiscent of the familiar physics in the spin chains as first discovered by Haldane\cite{haldane1983continuum}.

Our result also holds if we deform the Kitaev honeycomb model by other local interactions that commute with the local plaquette operators. 
One such interaction is the single ion anisotropy, $H_{SIA}=D_z\sum_i (S_i^z)^2$, which is suggested to be relevant to candidate Kitaev materials~\cite{xu2018interplay,xu2020possible}. In the presence of the magnetic field, time reversal symmetry is broken and possible Majorana cones are gapped out. This results in a non-Abelian topological order with Ising anyons as topological excitations. Its detailed nature is unclear and will be left for future study.

\emph{Acknowledgement.}-- 
HM would like to thank Liujun Zou for insightful discussions. 
Research at Perimeter Institute is supported in part by the Government of Canada through the Department of Innovation, Science and Economic Development Canada and by the Province of Ontario through the Ministry of Colleges and Universities.

\bibliography{ref}

\clearpage
\onecolumngrid

\setcounter{equation}{0}
\setcounter{figure}{0}
\setcounter{page}{1}
\renewcommand{\theequation}{S.\arabic{equation}}

\begin{center}
{\bf \large Supplementary Material of ``$\mathbbm{Z}_2$ spin liquids in the higher spin Kitaev honeycomb model:  an exact $\mathbbm{Z}_2$ gauge structure in a non-integrable model"}
\end{center}

\section{I. degenerate perturbation in the anisotropic limit}

As we discussed in the main text, in the anisotropic limit where $J_z \gg J_{x,y}$, there are extensive ground-state degeneracy. The ground states have spin configuration in the $Z$ basis satisfying $\langle S^z_i \rangle=\langle S^z_{\bar{i}}\rangle=\pm S$ where $i$ and $\bar{i}$ are two sites at the ends of a $z$ bond. This is given by 
the operator identity $S^z_i=S^z_{\bar{i}}$. If we decompose the spin-$S$ operator into $2S$ spin-$1/2$ operators, we obtain $(2S)^2$ operator identities $S^z_{a,i}=S^z_{b,\bar{i}}$ for $a,b\in [1,2S]$. If we further represent the spin-$S$ operator in terms of $8S$ Majorana fermions, they are constrained by relations $\gamma^0_{a,i}\gamma^z_{a,i}\gamma^z_{b,\bar{i}}\gamma^0_{b,\bar{i}}=1$ for any $a,b$. 

Within the space of the ground states, the effective Hamiltonian characterizes the fluctuation induced by small $J_x$ and $J_y$ terms. It can be obtained by the standard degenerate perturbation. Below, we discuss the effective Hamiltonian for the system of half-integer spins and integer spins, respectively.

\subsection{A. Half-integer spin system}

\begin{figure}[h]
    \centering
    \includegraphics[width=.48\textwidth]{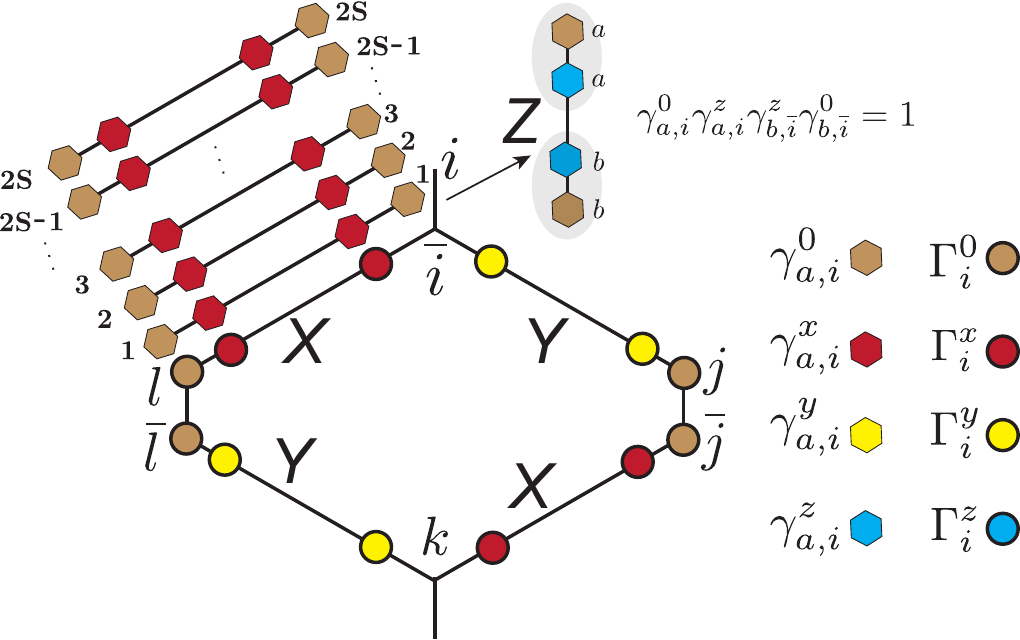}
    \caption{The half-integer spin Kitaev model reduces to the Wen plaquette model in the $J_z \gg J_x,J_y$ limit. As $J_z \rightarrow \infty$, the four Majorana fermions on the sites $i$ and $\bar{i}$ on the $z$ bond satisfy $S^z_{a,i}S^z_{b,\bar{i}}=\gamma^0_{a,i} \gamma^z_{a,i}\gamma^z_{b,\bar{i}}\gamma^0_{b,\bar{i}}=1$ for any $a,b\in[1,2S]$. Each plaquette term in Eq.~\eqref{eq:WP_plaquette} is a product of $4S$ terms on the $x$ bonds and $4S$ terms on the $y$ bonds of a hexagonal plaquette, giving a product of $32S$ Majorana partons ($\varhexagon$). It is equivalently a product of $12$ giant Majorana fermions ($\bigcirc$) around a plaquette.}
    \label{fig:Wen_plaquette}
\end{figure}

In the system of half-integer spins, we can get a non-trivial effective Hamiltonian at the $8S$-th order of the perturbation, given by
\begin{eqnarray}
   H_{\rm eff}^{\mathbbm{Z}+1/2} = -J_{\rm eff} \sum_{\diamond}  \mathcal{W}(\diamond) \label{eq:H_WP}
\end{eqnarray} 
with $J_{\rm eff} \propto J_x^{4S}J_y^{4S}/J_z^{8S-1} $. Each plaquette term can be obtained by a product of $4S$ terms on $x$ bonds and $4S$ terms on $y$ bonds given by
\begin{eqnarray}
  \mathcal{W}_p(\diamond) &=& \prod_{a=1}^{2S}S^x_{a,l} S^x_{a,\bar{i}} \prod_{b=1}^{2S}S^y_{b,\bar{i}}S^y_{b,j}
  \prod_{c=1}^{2S}S^x_{c,\bar{j}}S^x_{c,k} \prod_{d=1}^{2S}S^y_{d,k}S^y_{d,\bar{l}} \nonumber\\ 
  &=& \prod_{a=1}^{2S}\gamma^0_{a,\bar{i}}\gamma^x_{a,\bar{i}}\gamma^x_{a,l}\gamma^0_{a,l} \prod_{b=1}^{2S}\gamma^0_{b,\bar{i}}\gamma^y_{b,\bar{i}}\gamma^y_{b,j}\gamma^0_{b,j}\nonumber\\
  &\times& \prod_{c=1}^{2S}\gamma^0_{c,\bar{j}}\gamma^x_{c,\bar{j}}\gamma^x_{c,k}\gamma^0_{c,k}\prod_{d=1}^{2S}\gamma^0_{d,k}\gamma^y_{d,k}\gamma^y_{d,\bar{l}}\gamma^0_{d,\bar{l}} \nonumber\\
  &=& \Gamma_{\bar{i}}^y \Gamma_{j}^y \Gamma^0_{j} \Gamma^0_{\bar{j}} \Gamma_{\bar{j}}^x\Gamma_{k}^x \Gamma_{k}^y \Gamma_{\bar{l}}^y \Gamma^0_{\bar{l}} \Gamma^0_{{l}}\Gamma_{l}^x \Gamma_{\bar{i}}^x  \nonumber\\
   &=& \Gamma_{\bar{i}}^y \Gamma_{j}^y \Gamma^z_{j} \Gamma^z_{\bar{j}} \Gamma_{\bar{j}}^x\Gamma_{k}^x \Gamma_{k}^y \Gamma_{\bar{l}}^y \Gamma^z_{\bar{l}} \Gamma^z_{{l}}\Gamma_{l}^x \Gamma_{\bar{i}}^x  .
  \label{eq:WP_plaquette}
\end{eqnarray}
In the first line, the degenerate perturbation is done in terms of spin-$1/2$ operators $S^{x,y}_{a,i}$. Due to the constraints $S^z_{a,i }=S^z_{b,i }$ for any $a,b \in [1,2S]$, on site $i$ we can define $S^z_i=S^z_{a,i}$ whose conjugate operator is $\prod_{a=1}^{2S}S^{x(y)}_{a,i}$. This calls for a product of terms $J_xS^{x(y)}_{a,i}S^{x(y)}_{a,i+x(y)}$ with $2S$ different flavor $a$ in the effective Hamiltonian. Put it in another way, in the basis of $S^z$, all the $2S$ spin-$1/2$ on the same site should be flipped together by acting all $S^{x(y)}_{a,i}$ simultaneously.
In the second line, each plaquette term is a product of Majorana fermions and in the third line this expression is simplified using giant parton operators on the bonds around the plaquette, as shown in Fig.~\ref{fig:Wen_plaquette}. 
Notice $\mathcal{W}(\diamond)$ is exactly the flux terms $W_p$ in Eq.~\eqref{eq:flux_hexagon}.

\begin{figure}[h]
    \centering
    \includegraphics[width=.8\textwidth]{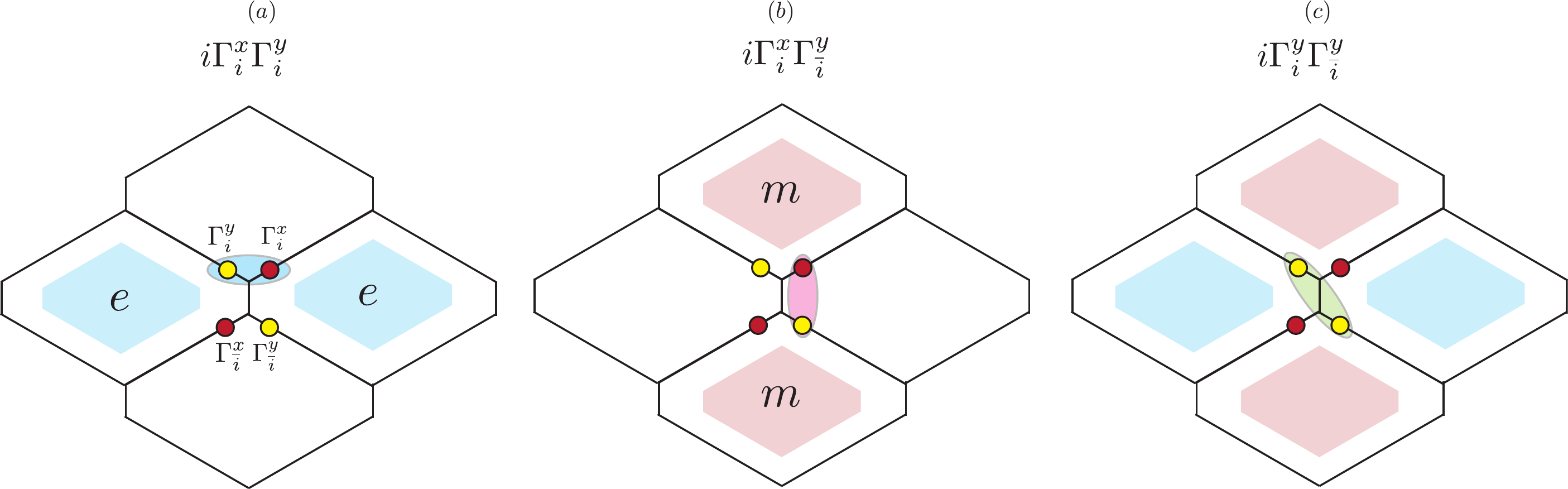}
    \caption{The $e$ (a), $m$ (b) and $\varepsilon$ (c) excitations of the Wen plaquette model are created by acting the giant Majorana fermions on the ground states. }
    \label{fig:Wen_plaquette_excitations}
\end{figure}

This effective Hamiltonian is exactly the Wen-Plaquette model. It has $\mathbbm{Z}_2$ topological order. Recall the constraint $\Gamma^0_{i}\Gamma^z_i\Gamma^z_{\bar{i}}\Gamma^0_{\bar{i}}=1$. Only states with an even number of giant Majorana fermions on each site and on any $z$ bond are physical.
In Fig.~\ref{fig:Wen_plaquette_excitations}, we list all the possible physical topological excitations.

\subsection{B. Integer spin system}

\begin{figure}[h]
    \centering
    \includegraphics[width=.48\textwidth]{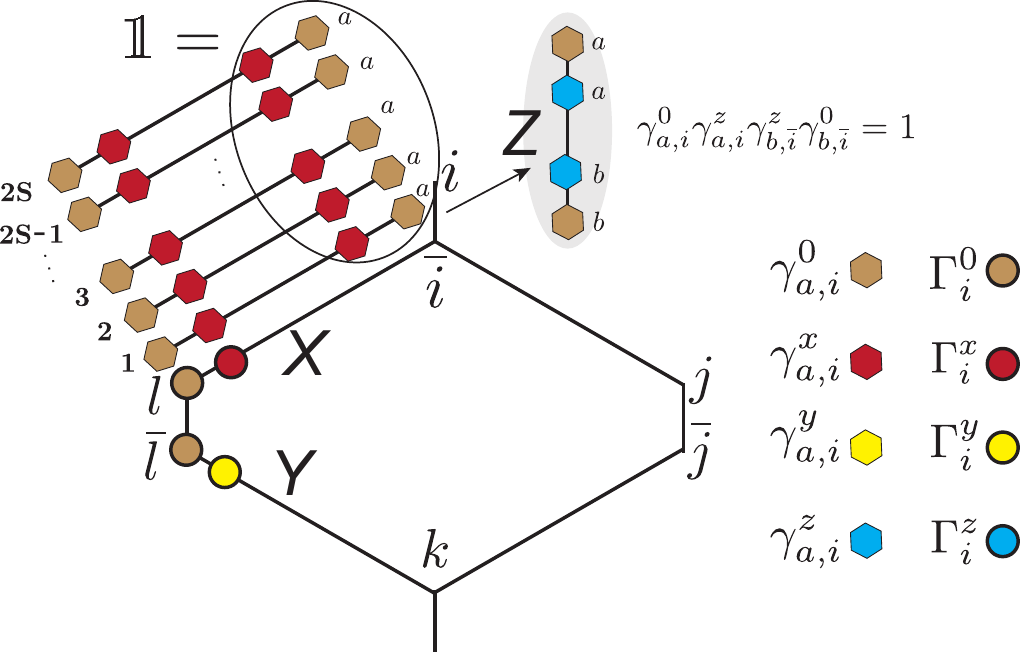}
    \caption{The effective Hamiltonian of the integer spin Kitaev model in the anisotropic limit is non-trivial at the $4S$-th order, which is contributed by the $4S$ terms on the $x$ and $y$ bonds attaching to a $z$ bond. }
    \label{fig:integer_spin_Heff}
\end{figure}

The effective Hamiltonian for a system of integer spins is different. Up to the $4S$-th order in the perturbation theory, we get a non-trivial Hamiltonian
\begin{eqnarray}
     H_{\rm eff}^{\mathbbm{Z}} &=& -J_{\rm eff} \sum_{i} \prod_{a=1}^{2S}S^x_{a,i}(S^x_{c,i+x})^{2S}\prod_{b=1}^{2S}S^y_{b,\bar{i}}(S^y_{d,\bar{i}+x})^{2S} \nonumber\\
     &=&-J_{\rm eff} \sum_{i}  (\gamma^0_{c,i+x} \gamma^x_{c,i+x})^{2S} \prod_{a=1}^{2S} \gamma^x_{a,{i}} \gamma^0_{a,{i}} \prod_{b=1}^{2S} \gamma^0_{b,\bar{i}} \gamma^y_{b,\bar{i}} (\gamma^y_{d,\bar{i}+x} \gamma^0_{d,\bar{i}+x})^{2S} \nonumber\\
   &=& -J_{\rm eff} \sum_{i} \Gamma^0_{i}\Gamma^x_{i}\Gamma^y_{i+z}\Gamma^0_{i+z}=-J_{\rm eff} \sum_{i}  ~e^{i\pi S^x_{i}} e^{i\pi S^y_{i+z}} .
\end{eqnarray}
In the first line, in terms of the spin-$1/2$ operators, we take product of terms $J_{x(y)} S^{x(y)}_{a,i}S^{x(y)}_{c,i+x(y)}$ for all $a\in [1,2S]$ but fixed $c$ as shown in Fig.~\ref{fig:integer_spin_Heff}. This actually cancels any non-trivial operator on the site $i+x(y)$ since $(S^{x(y)}_{a,i})^{2S}=\mathbbm{1}$ for even $2S$. Accordingly, the effective Hamiltonian only contains ultra-local terms of the giant partons on the $z$ bonds. Furthermore, since $(e^{i\pi S^x_{i}} e^{i\pi S^y_{i+z}}  )^2=1$, we can write this bosonic operator as a Pauli matrix, i.e. $e^{i\pi S^x_{i}} e^{i\pi S^y_{i+z}} =\sigma^x$. Then the effective Hamiltonian has a unique featureless ground state as a trivial paramagnet, i.e. $\sigma^x=1$.

\section{II. USp(4S) parton construction \label{app:USp}}

In fact, the construction in Eq.~\eqref{eq:parton} is a $USp(4S)$ parton construction. For general spin-$S$, we can write
\begin{eqnarray}
    S^\mu = -\frac{1}{4} \textrm{Tr} (F^\dag F\sigma^\mu),
    \label{eq:parton_USP}
\end{eqnarray}
where 
\begin{eqnarray}
    F=\left[\begin{array}{ccccccc}
       f^\dag_1 &\dots & f^\dag_{2S}  & & f^\dag_{2S+1}& \dots & f^\dag_{4S}  \\
       f_{2S+1} &\dots & f_{4S}  & & -f_{1}& \dots & -f_{2S}
    \end{array}\right]^T.
    \label{eq:fermion_matrix}
\end{eqnarray}
More concretely, we can write 
\begin{eqnarray}
    S^x &=& -\frac{1}{2} \sum_{a=1}^{2S} (f_a f_{2S+a}+ f_{2S+a}^\dag f_a^\dag ), \nonumber\\
    S^y &=& -\frac{i}{2} \sum_{a=1}^{2S} (f_a f_{2S+a}- f^\dag_{2S+a}  f_a^\dag ) ,\nonumber\\
    S^z &=& -\frac{1}{2} \sum_{a=1}^{4S} (f_a f_a^\dag -\frac{1}{2}).
\end{eqnarray}
The $4S$ fermions in the matrix $F$ in Eq.~\eqref{eq:fermion_matrix} satisfy local constraint
\begin{eqnarray}
    F \left[\begin{array}{cc}
       0  & 1 \\
        -1 & 0
    \end{array}\right] F^T =
    \left[\begin{array}{cc}
       0  & -I_{2S} \\
        I_{2S} & 0
    \end{array}\right]. \label{eq:local_constraint}
\end{eqnarray}
This gives equations
\begin{eqnarray}
     f^\dag_a f_{2S+b} &=& f_{2S+a}f_b^\dag, \quad \quad f_{a}f_{2S+b}^\dag  = f^\dag_{2S+a} f_{b},\nonumber\\
    f_{2S+a}f_{2S+b}^\dag +f^\dag_a f_{b} &=& \delta_{ab}, \quad \quad 
    f_{a}f_b^\dag +f^\dag_{2S+a} f_{2S+b} =\delta_{ab} .
    \label{eq:constraint_1}
\end{eqnarray}
With the anti-commutation relations of the fermions $f^\dag_a f_b+f_bf^\dag_a =\delta_{ab}$, we can get $f^\dag_a f_{a} = f_{2S+a}^\dag f_{2S+a} $. With Eq.~\eqref{eq:constraint_1},
one can check the spin operators satisfy $[S^\mu, S^\nu]=i \varepsilon^{\mu\nu\lambda}S^\lambda$ and also $\sum_\mu (S^\mu)^2= S(S+1)$. 

When $S=1/2$, $USp(2)\approx SU(2)$. This is the familiar $SU(2)$ Schwinger fermion parton construction where we can write
$
    S^\mu =\frac{1}{2} f^\dag \sigma^\mu f 
   $
where $f=(f_\uparrow  , f_\downarrow^\dag )^T$. Namely
\begin{eqnarray}
    S^x &=&-\frac{1}{2} (f_\uparrow f_\downarrow +f_\downarrow^\dag f_\uparrow^\dag), \nonumber\\
    S^y &=& -\frac{i}{2} (f_\uparrow f_\downarrow -f_\downarrow^\dag f_\uparrow^\dag), \nonumber\\
    S^z &=& -\frac{1}{2} (f_\uparrow f_\uparrow^\dag +f_\downarrow f_\downarrow^\dag -1 ).
\end{eqnarray}
And the fermionic operators satisfy relations 
\begin{eqnarray}
    f^\dag_{\uparrow}f_\uparrow + f_{\downarrow}f_\downarrow^\dag=1.
    \label{eq:constraint_SU(2)}
\end{eqnarray}
If we further write $f_\uparrow^\dag= \frac{1}{2}(\gamma^0-i\gamma^z)$ and $f_\downarrow^\dag = \frac{1}{2}(-\gamma^y+i\gamma^x)$, we can reproduce the four Majorana representation for spin-$1/2$. Firstly, the Eq.~\eqref{eq:constraint_SU(2)} is equivalent to $\gamma^0\gamma^x\gamma^y\gamma^z=1$. Secondly, we can reproduce the operator identities $2S^\mu=i\gamma^0\gamma^\mu$.

Therefore, for general $S$, we can write $f^\dag_a=\frac{1}{2}(\gamma^0_a-i\gamma^z_a)$ and $f^\dag_{2S+a}=\frac{1}{2}(-\gamma^y_a+i\gamma^x_a)$. This gives the local constraints $\gamma^0_a\gamma^x_a\gamma^y_a\gamma^z_a=1$ and the operator identities $2S^\mu=i\sum_{a=1}^{2S} \gamma^0_a \gamma^\mu_a$, which is exactly Eq.~\eqref{eq:parton}. The other constraints in Eq.~\eqref{eq:constraint_1} can be reorganized as $\sum_\mu (\sum_{a=1}^{2S} \gamma^0_a \gamma^\mu_a)^2= - S(S+1)$.

\end{document}